\documentclass[a4paper,11pt]{article}

\usepackage{jcappub} 

\usepackage{graphicx}
\usepackage[dvipsnames]{xcolor}
\usepackage{soul}
\usepackage{amstext}
\usepackage{bm}				
\usepackage{booktabs}		
\usepackage{cleveref}		
\usepackage[section]{placeins}	
\usepackage{pgfplots}		
\usepackage{tabularx}
\usepackage{multirow}
\usepackage{arydshln}

\usepackage{caption}
\usepackage[caption=false]{subfig}
\usepackage{float}
\usepackage{subfloat}
\newcolumntype{H}{>{\centering\arraybackslash}X}

\newcommand{\be}{\begin{equation}}
\newcommand{\ee}{\end{equation}}
\newcommand{\ba}{\begin{eqnarray}}
\newcommand{\ea}{\end{eqnarray}}

\title{Toward a direct measurement of the cosmic acceleration: roadmap and forecast on FAST}

\author[a]{Kang Jiao}
\author[a,b,c]{, Jian-Chen Zhang} 
\author[a,b]{, Tong-Jie Zhang}
\author[d]{, Hao-Ran Yu}
\author[e]{, Ming Zhu}
\author[f,g]{and Di Li} 

\affiliation[a]{Department of Astronomy, Beijing Normal University, Beijing 100875, China}
\affiliation[b]{Institute for Astronomical Science, Dezhou University, Dezhou 253023, China}
\affiliation[c]{College of Automotive Engineering, Dezhou University, Dezhou 253023, China}
\affiliation[d]{Department of Astronomy, Xiamen University, Xiamen, Fujian 361005, China}
\affiliation[e]{National Astronomical Observatories, Chinese Academy of Sciences, Beijing 100012, China}
\affiliation[f]{CAS Key Laboratory of FAST, NAOC, Chinese Academy of Sciences, Beijing 100101, China}
\affiliation[g]{University of Chinese Academy of Sciences, Beijing 100049, China}

\emailAdd{tjzhang@bnu.edu.cn} 

\date{\today}

\abstract{
HI absorption systems are great targets for direct measurement of the Sandage-Loeb (SL) effect throughout a wide range of redshift for ground-based radio telescopes. We demonstrate the significance of improving the accuracy of SL effect measurement in cosmological model selection. With its wide sky coverage and high sensitivity, we forecast that for 1 year of the upcoming commensal survey (CRAFTS) the FAST telescope is capable of discovering about 800 HI absorption systems thereby improving the SL measurement accuracy.  Aiming to measurement the cosmic redshift drift rate at the precision of $\dot{z} \sim 10^{-10} \mathrm{decade^{-1}}$, we propose combined observation mode with blind-searching and targeted observation. For a decade of consecutive targeted spectroscopic observation with the frequency resolution at a level of sub-$0.1\ \rm Hz$, we could detect the first-order derivative of the cosmological redshift with the expected precision.
}

\begin{document}

\maketitle
\flushbottom

\section{Introduction}\label{sec:introduction}
The concept that our universe is currently undergoing an accelerating expansion stage has been strongly confirmed after its discovery by type Ia supernova (SN Ia) observations \cite{Riess:1998, Perlmutter:1999}. Dark energy is the most accepted explanation that could driven this accelerating. But as to its equation of state (EoS) parameter $\omega_{\rm de}\equiv p_{\rm de}/\rho_{\rm de}$, current observations have not precisely distinguished between the cosmological constant $\Lambda$ and the evolved $\omega(z)$. Today, cosmological probes which include Cosmic Microwave Background (CMB) \citep{Planck:2015, Planck:2018}, type Ia supernovae as  standard candles, baryon acoustic oscillations (BAO) from galaxy clusterings as standard rulers \citep{Eisenstein:2005},  gravitational waves as standard sirens \citep{Chen:2018} and so on are kept advancing the study about the composition and dynamical evolution of the universe.

Considering that all these probes observe the cosmological dynamics $\ddot{a}$ indirectly, so it is of great importance to measure the velocity change of celestial bodies moving passively with the Hubble flow in real-time. This basic idea first became feasible at the time of Sandage proposed to measure the change of redshift of galaxies \cite{Sandage:1962}, and was advanced to measure such change in the Ly$\alpha$ forest by Leob \cite{Loeb:1998}. This phenomenon is the well-known SL effect
and also referred to as the redshift drift. The SL effect provides us an as independent cosmological probe which can be applied in various researches, including the dark energy redshift desert \citep{Corasaniti:2007}, constraints of other dark energy models \citep{Balbi:2007,Martinelli:2012,Zhang:2010} and modified gravity theories \citep{Jain:2010,Li:2013}. Yuan \& Zhang \cite{Yuan:2015} proposed a robust scheme to measure the high redshift ($2.0\le z \le 5.0$) Hubble parameters $H(z)$ and to reduce the $\Omega_\Lambda\text{-}\Omega_{\rm m}$ and $H_0\text{-}\Omega_{\rm m}$  degeneracy for CODEX (COsmic Dynamics and EXo-earth experiment)-like survey. Melia \cite{Melia:2016} discussed the possibilities in discriminating cosmological models such as $R_h=ct$ model and $\Lambda \rm CDM$ model. 

Darling \cite{Darling:2012} re-observed the known HI absorption systems over more than a decade using the Green Bank Telescope (GBT) and  got the best measurement on the SL effect. Although these measurements are three orders of magnitude larger than the theoretically expected values, they demonstrate the lack of the peculiar acceleration in absorption line systems and the long-term frequency stability of modern radio telescope. Therefore even more precise measurement can be expected by expanding the sample, using longer time baselines, larger redshifts, and higher signal-to-noise observations, particularly from larger apertures. Yu et al.~\cite{Yu:2014} forecasted that for a CHIME-like HI absorption systems survey with a decadal time span, we can detect the acceleration of a $\Lambda$CDM universe with 5$\sigma$ confidence.  Benefiting from its high sensitivity and wide field, the Five-hundred-meter Aperture Spherical radio Telescope (FAST) \citep{Nan:2011} with the 19-beam system will have a great advantage in blind-searching the 21-cm hydrogen absorption systems. 

In this paper, we mainly propose a observational roadmap and forecast the prospect of FAST in the SL effect measurement. In section \ref{sec:principle}, we distinguish the physical meaning between three different definitions of acceleration that are associated with the expansion of the universe. Then we exam the dependence of the $\dot{z}$ on cosmological models to show its potential in model selection. In section \ref{sec:parkes}, through the comparison of the PKS B1740-517 spectrum from two distinct telescope observations, we emphasize the necessity of a long-term consecutive high-resolution spectroscopic observation. In section \ref{sec:FAST}, we propose a blind-searching plus targeted-observation combined observation mode for SL measurement and predict the performance of the Commensal Radio Astronomy FAST Survey (CRAFTS) in such observation. In section \ref{sec:conclu}, we draw our conclusions and make some discussions.  

\section{SL effect in cosmology}\label{sec:principle}

The cosmological redshift is the most readily observable evidence of the expansion of the universe. For a fixed comoving distance, the redshift $z$ of a celestial body changes with time due to the accelerating expansion of the universe, which is the  well-known Sandage-Loeb (SL) effect and also referred to as the redshift drift. The average change rate of a source's redshift is approximately
\begin{align}
\frac{\Delta z}{\Delta t_0}\approx\frac{\dot{a}(t_0)-\dot{a}(t_{\rm s})}{a(t_{\rm s})},\label{Eq:Dotz}
\end{align}
where $a$ is the scale factor, and $t_0$ is the observation time while $t_{\rm s}$ is the radiation emission time.  Taking the limit approximation of Eq.~\eqref{Eq:Dotz} as the first-order derivative of redshift respect to the current cosmic time, which is theoretically determined as
\begin{equation}
\dot{z}=\frac{{\rm d}z}{{\rm d}t_0} = H_0[1+z]-H(z),
\end{equation} where $H(z)=\dot{a}/a$ is Hubble parameter (or expansion rate) and $H_0$ is its current value named as Hubble constant. $H(z)$ is related to the specific recipe of the cosmic components as 
\begin{align}
H(z)=H_0\sqrt{\sum_{i}\Omega_{i}(1+z)^{3(1+\omega_i)}}\label{Equ:E(z)},
\end{align}
where $i$ denotes dark energy, space curvature, matter, and radiation, while $\Omega_i$ is the density parameter and $\omega_i$ is the EoS parameter.

\begin{figure}
	\centering
	\includegraphics[width=0.8\textwidth]{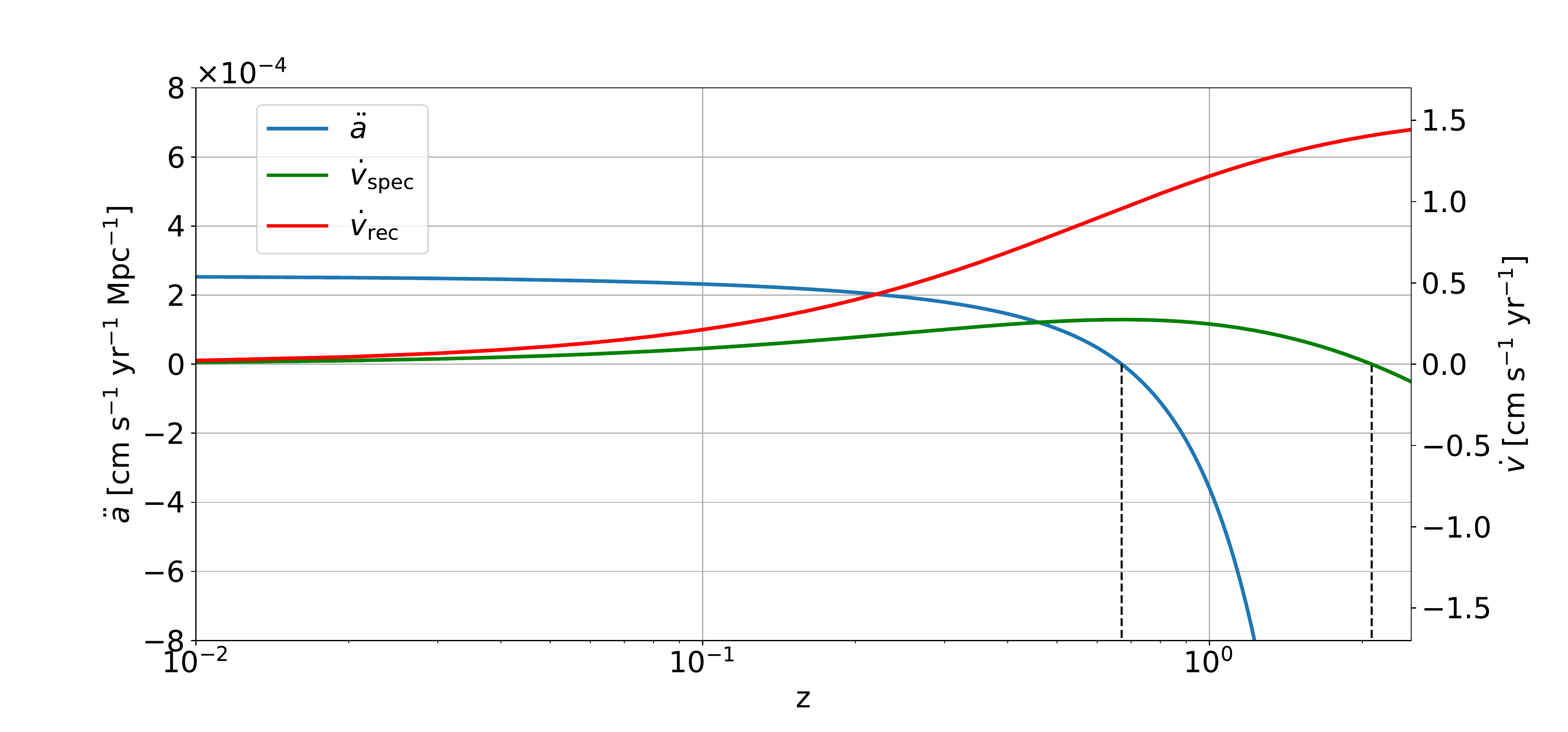}
	\caption{Theoretical cosmic acceleration $\ddot{a}$ (blue), the spectroscopic acceleration $\dot{v}_{\mathrm{spec}}$ (green) and the recession acceleration $\dot{v}_{\rm rec}$ (red) respectively for a fiducial model with $\Omega_{m0}=0.3,\ \Omega_{\rm de0}=0.7$ and $\omega_{\rm de} = -1$. Ordinate value greater than zero means accelerating, and the dotted vertical lines mark out the redshifts that indicate the decelerating turns to accelerating.
	}
	\label{fig:def}
\end{figure}

At low redshift, a comoving source's spectroscopic speed shift $\Delta v_{\mathrm{spec}}$ over an observing period $\Delta t_0$ is usually taken as its average recession acceleration  (e.g. in Ref.~\cite{Loeb:1998,Yu:2014,Melia:2016})
\begin{equation}
\dot{v}_{\mathrm{spec}}=\frac{\Delta v_{\mathrm{spec}}}{\Delta t_0}=\frac{c}{1+z}\frac{\Delta z}{\Delta t_0}.\label{Eq:dotv}
\end{equation}
However, the velocity due to the rate of expansion of space cannot be calculated through the special relativistic (SR) Doppler effect.  We need to interprate the cosmic expansion in a general relativistic (GR) way, in which the relation between redshift and recession velocity should be \cite{Davis:2001,Davis:2004}
\begin{align}
	v_{\mathrm{rec}}(t,z) = \frac{c}{a_0}\dot{a}(t)\int_0^z\frac{\mathrm{d}z^\prime}{H(z^\prime)}.
\end{align} 
Differentiating this yields the recession acceleration we observed to be
\begin{align}
	\dot{v}_{\rm rec}(z) = \ddot{a}_0\chi(z) +\dot{a}_0\frac{{\rm d}\chi(z)}{{\rm d}z}\frac{{\rm d}z}{{\rm d}t_0},
\end{align}
where $\chi(z) = c\int_0^z\frac{\mathrm{d}z^\prime}{H(z^\prime)}$ is the comoving distance. 
The Eq.~\eqref{Eq:dotv} is workable only because the low redshift approximation of velocity are the same in both the SR and GR results (as shown in Figure~\ref{fig:def}).
Further discussions about the most common misconceptions of the cosmic expansion can be found in Ref.~\cite{Davis:2004}.
Besides, the spectroscopic acceleration $\dot{v}_{\mathrm{spec}}$ is also in different with the cosmic acceleration $\ddot{a}$ which can be calculated from the Friedmann's second equation 
\begin{align}
\ddot{a} = -\frac{aH^2}{2}\sum_{i}\Omega_i(1+3\omega_i).
\end{align}
To clarify the differences between them, we calculate $\ddot{a}$, $\dot{v}_{\mathrm{spec}}$ and $\dot{v}_{\mathrm{rec}}$ within a fiducial model with $\Omega_{m0}=0.3,\ \Omega_{\rm de0}=0.7$ and $\omega_{\rm de} = -1$. As shown in Fig.~\ref{fig:def}, the cosmic acceleration $\ddot{a} >0$ when $z\lesssim0.67$, while the spectroscopic acceleration $\dot{v}_{\mathrm{spec}} >0$ when $z\lesssim2.09$.  It should be pointed out that the expansion of the universe is said to be ``accelerating" if $\ddot{a} > 0 $, but not in the case of $\dot{v}_{\mathrm{spec}}$ be negative. In summary, it is more reasonable to use the $\dot{z}$ than $\dot{v}_{\mathrm{spec}}$ to quantify the SL effect, especially in high redshift.

\begin{figure}
	\centering
	\includegraphics[width=0.7\textwidth]{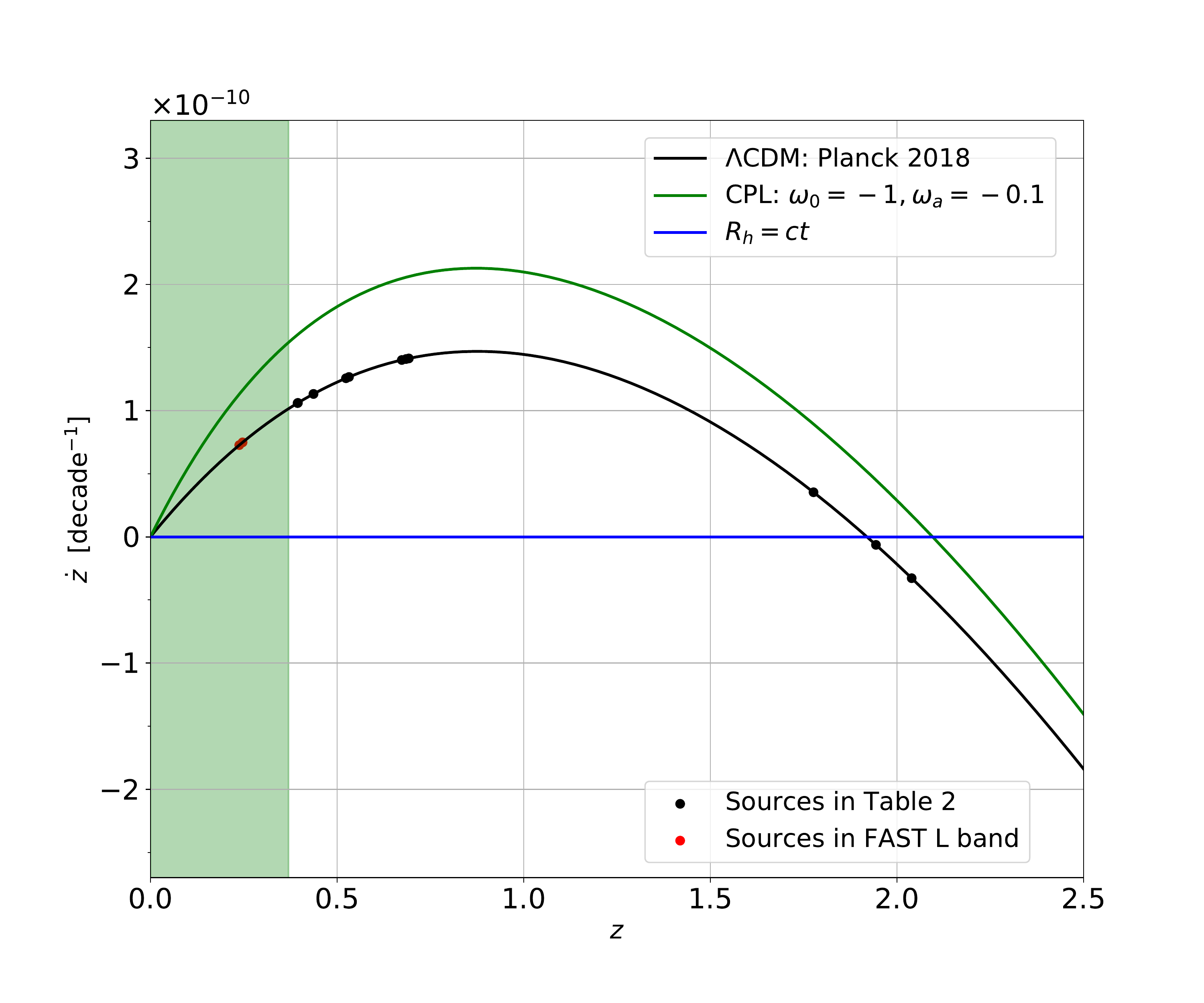}
	\caption{The theoretical change rate of the cosmic redshift of a comoving object (Eq.~\eqref{Eq:Dotz}) as predicted by FRW cosmology for  $\Lambda \rm CDM$ model with the {\slshape{Planck 2018}} result (black), CPL model with $\omega_{0}=-1,\omega_{a}=0.1$ (green) and $R_h=ct$ model (blue). The horizontal axis represents the initial redshifts and the vertical axis is the change rate of cosmic redshift in a unit of per decade.  The green shaded region shows the observable redshift range covered by the FAST 19-beam L band receiver. The scattered vertical lines mark out the redshifts of sources to be measured in Table \ref{tab:sources}.
	}
	\label{fig:calculation}
\end{figure}

The model-independent measurement of $\dot{z}$ makes it a powerful probe in cosmological model discrimination. We compare the $\dot{z}$ between the concordance $\Lambda {\rm CDM}$ model, Chevallier-Polarski-Linder (CPL) parametrization \cite{Chevallier:2001,Linder:2003} in which $\omega_{\rm de} = \omega_{0}+\omega_{a}z/(1+z)$, and $R_h=ct$ model \citep{Melia:2013} as examples here. We choose the best fit parameters of flat-$\Lambda {\rm CDM}$ from \cite{Planck:2018}, where $H_0 = 67.36\ {\rm km\ s^{-1}\ Mpc^{-1}}$, $\Omega_{de} = 0.685$, $\Omega_m = 0.315$,  $\Omega_{k} = 0$, $\omega_{de} = -1$. For CPL model, we set parameters the same as the former but $\omega_{0}=-1,\omega_{a}=0.1$. And for $R_h=ct$ model, the hubble parameter $H(z)=H_0[1+z]$, thus without redshift drift \citep{Melia:2016}. Setting the parameters as above, we can get the acceleration-redshift relation for each model shown in Fig.~\ref{fig:calculation}. As we can see, if we could measure the $\dot{z}$ at the precision level of $\sim 10^{-10} \ \rm decade^{-1}$, these models can be distinguished with a great significance. However, an extreme high spectroscopic resolution and long-term observation is required to fulfill this ambition. This issue is specified in section~\ref{sec:FAST}.

\section{Test observation and plan for FAST}\label{sec:observation}
Hydrogen is the most ancient and abundant element in the universe, that makes it a promising indicator to investigate the evolutionary history of our universe throughout a wide redshift range. Benefit from the narrow intrinsic line width of 21 cm HI absorption, which is dominated by cold absorbers at $T<80\ \rm K$,  the redshift drift in such a system could be measured by a long-term high-resolution spectroscopic ground base radio observation. 

\subsection{Observation with Parkes}\label{sec:parkes}

In 2014, Allison et al. \cite{Allison:2015} (hereafter as BETA14) discovered a new 21-cm HI absorption system in a blind-searching mode using the commissioning data from the Boolardy Engineering Test Array (BETA) \citep{Hotan:2014} of the Australian Square Kilometre Array Pathfinder (ASKAP) \citep{Johnston:2008}. The absorption line is detected at $z \sim 0.44$ towards the GHz-peaked spectrum radio source PKS B1740-517, which is centered on ${\rm RA(J2000)=17^h44^m25^s.45} $, $\rm DEC(J2000)=-51^\circ44^\prime43^{\prime\prime}.8$. 

Aiming to observe its redshift drift, we re-observed this source to compare the position of the peak absorption during the PX501 observing time with Parkes on July 1st, 2018. We adopt the source-on and source-off observation mode with the Ultra-wideband Low (UWL) receiver lasting for an hour duration. The only drawback is that an error of the UWL backend GPU cluster existed during the observation period, thus we only managed to record 32768 channels across each 128 MHz band that gives a channel spacing of 3.9 kHz resolution, that is $1.18\ \rm km/s$ at the spectral line position $ 985.5\ \rm MHz$. The bug in the GPU code is now fixed providing a promising 60 Hz resolution, which is equivalent as $18\ \rm m/s$ in velocity. 

We subtract the continuum spectral in a linear approximation, then we apply the \textsl{radial\_velocity\_correction}\footnote{http://docs.astropy.org/en/stable/coordinates/velocities.html}  method in \textsl{Python} module \textsl{astropy} \citep{Robitaille:2013} to do the solar barycentric correction on the spectrum. The correction formula is
\begin{align}
v_t=v_m+v_\mathrm{b}+\frac{v_\mathrm{b}v_\mathrm{m}}{c},
\end{align}
where $v_t$ is the true radial velocity, $v_m$ is the measured radial velocity, $v_b$ is the barycentric correction returned by \textsl{radial\_velocity\_correction} and $c$ is the speed of light in vacuum respectively. The equivalent measured radial velocity for radio frequency is 
\begin{align}
	v_\mathrm{m} = \frac{\nu_0-\nu_\mathrm{m}}{\nu_0}c,
\end{align}
where $\nu_0$ is the rest frequency of the 21 cm line and $\nu_\mathrm{m}$ is the measured frequency.
The correction accuracy is at the acceptable level of approximately $3\ \rm m/s$, which is almost up to 3 orders smaller than the spectrum resolution. We also double check the correction result using the \textsl{pyasl.helcorr} in the \textsl{PyAstronomy}\footnote{https://github.com/sczesla/PyAstronomy} module.

\begin{figure*}[htbp]
	\centering
	\subfloat[]{%
		\includegraphics[height=.2\textheight]{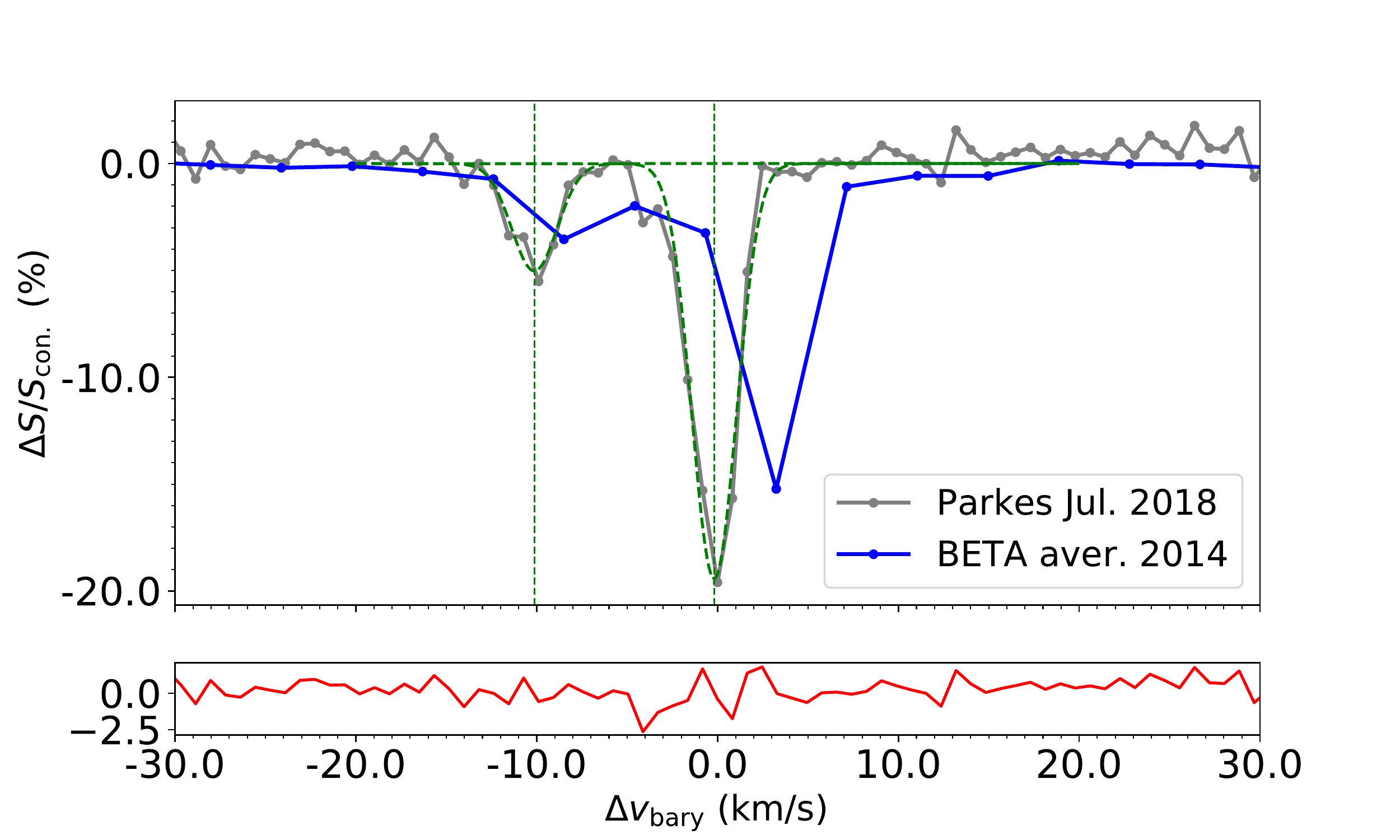}}\quad
	\subfloat[]{%
		\includegraphics[height=.2\textheight]{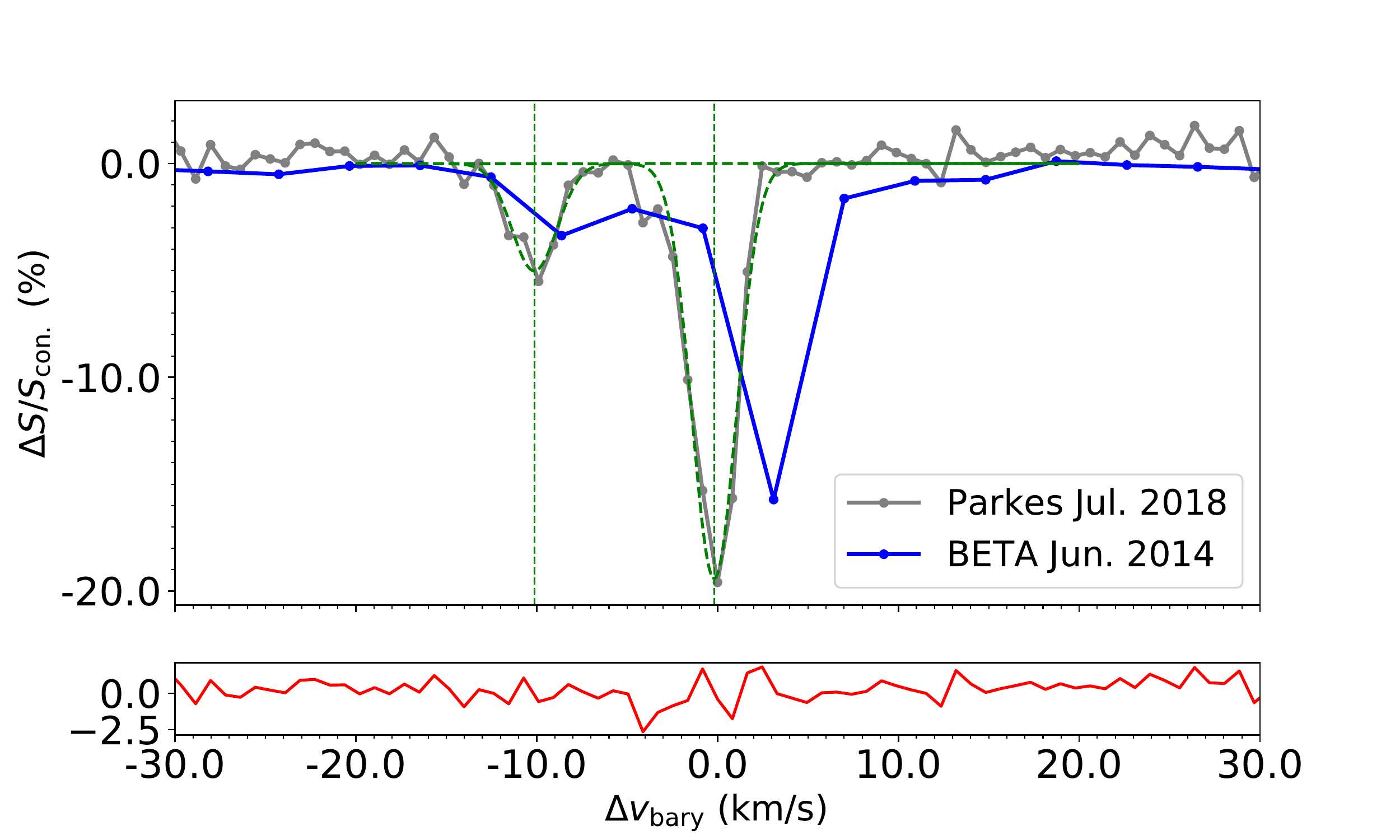}}\\
	\subfloat[]{%
		\includegraphics[height=.2\textheight]{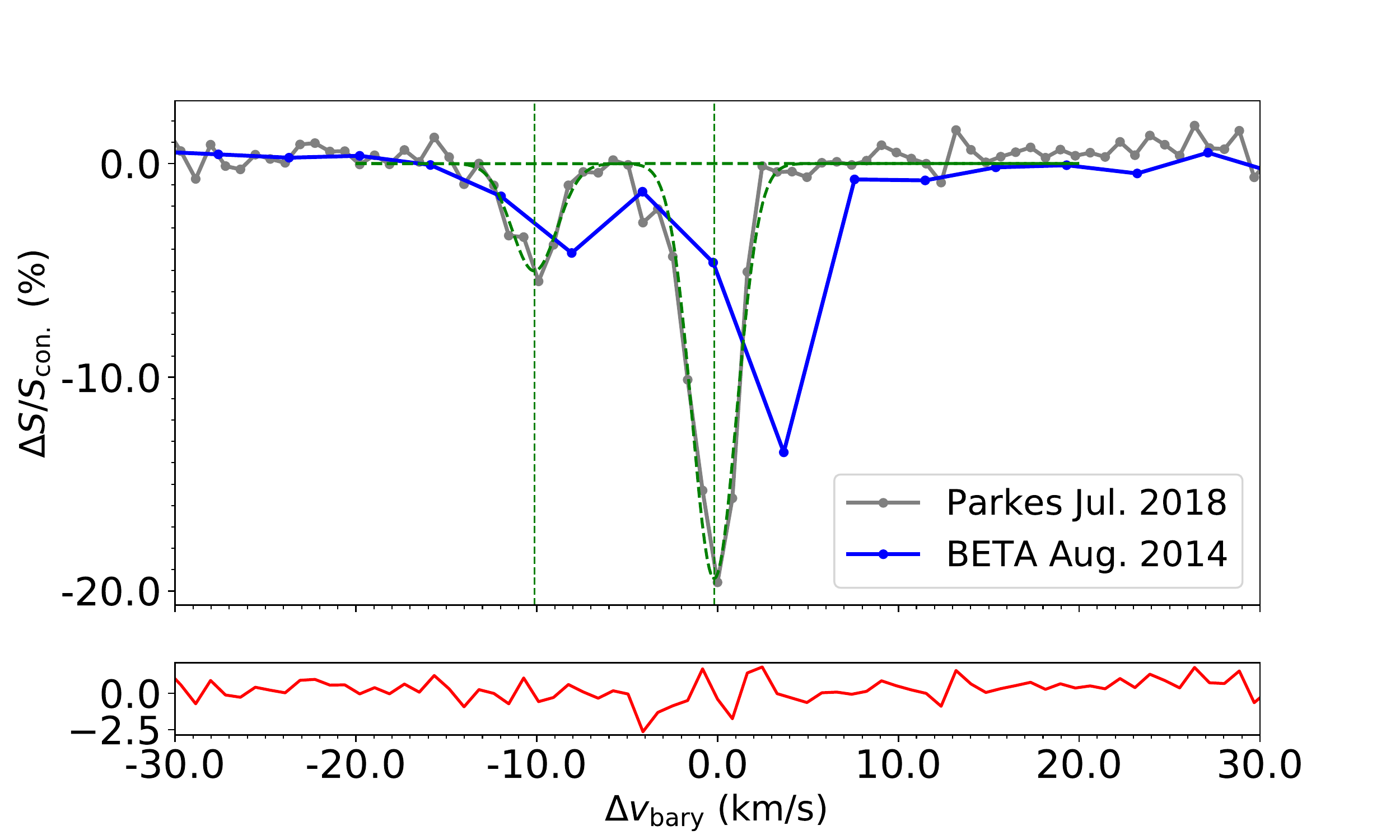}}\quad
	\subfloat[]{%
		\includegraphics[height=.2\textheight]{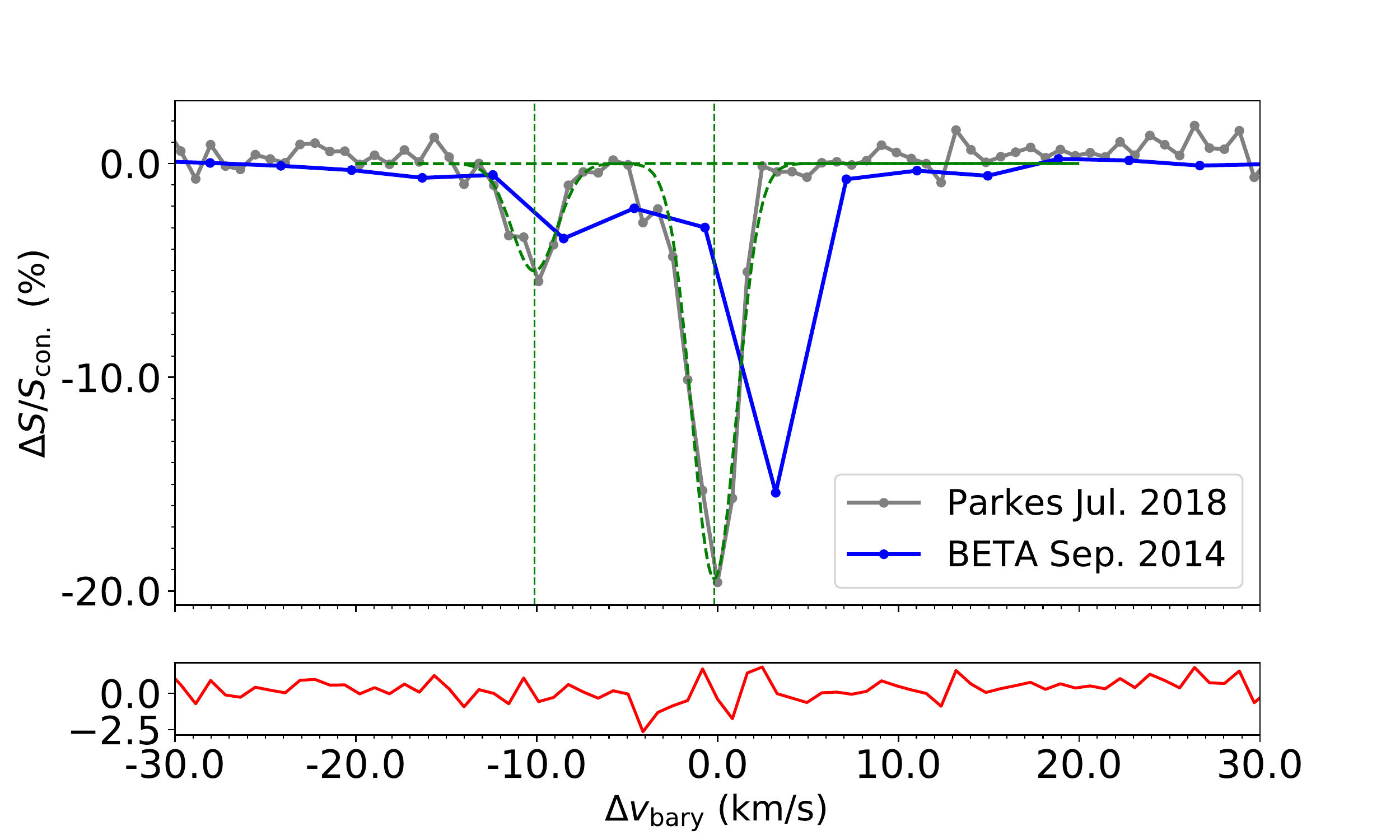}}\\
	\caption{A comparison of HI absorption spectrum for source PKS B1740-517. The radial velocity axis is given relative to the peak absorption position of which the rest frame is defined in our Parkes observation in 2018 (grey line). The green dashed spectral lines are the two gaussian components whose peak position are given by the vertical dashed lines. The blue line is the average spectrum of 3 BETA observations in 2014 \citep{Allison:2015}.  Obviously, an offset between the spectrum from Parkes and BETA exists within the velocity resolution of BEAT. The red solid line in each of bottom panels denote the best-fitting residual.}\label{fig:spetrum}
\end{figure*}

Comparing to the average HI absorption line from the BETA14 observation, an approximately $3\ \rm km/s$ blue shift exists between the two sets of observations shown as Table  \ref{tab:comparison} and Fig.~\ref{fig:spetrum}.  However, the pairwise difference of the radial velocity at the peak position is within the spectral resolution of the BETA, which is $18.5 \rm kHz$ (equivalent to $ 5.6\  \rm km/s$ at $ 985.50846\ \rm MHz$). Therefore, it is not sufficient to determine whether the two spectral lines are offset by $\sim 3\ \rm km/s$ or not, which is highly dependent on the definition of the line “position” (e.g. at peak optical depth or optical-depth weighted position etc.) or the model used for fitting the lines. We need at least one more Parkes observation to determine this. And for the same resolution consecutive observations in Ref.~\cite{Allison:2018}, no such obvious systematic line-shift has been observed.

\begin{table*}
	\centering\caption{A comparison of model parameters derived from fitting the HI absorption line seen towards PKS B1740−517 for each epoch and the average spectrum. }\label{tab:comparison}
	\setlength{\tabcolsep}{2mm}{
		\begin{tabular}{lllcll}
			\hline
			\multicolumn{1}{l}{Epoch} & \multicolumn{1}{c}{ID} & \multicolumn{1}{c}{$z_{\rm bary}$} & \multicolumn{1}{c}{$\Delta{v_{50}}$} &  \multicolumn{1}{c}{$(\Delta{S}/S_{\rm cont})_{\rm peak}$} & \multicolumn{1}{c}{$\Delta{v_{\rm peak}}$}\\
			\multicolumn{1}{c}{} & \multicolumn{1}{c}{} & \multicolumn{1}{c}{} & \multicolumn{1}{c}{(km\,s$^{-1}$)} & \multicolumn{1}{c}{(\%)} & \multicolumn{1}{c}{(km\,s$^{-1}$)} \\
			\hline
			2014 June 24 & 1 & $0.44129264_{-0.00000060}^{+0.00000058}$ & $5.15_{-0.21}^{+0.20}$ & $-20.20_{-0.74}^{+0.68}$ & $2.994$ \\
			& 2 & $0.4412223_{-0.0000022}^{+0.0000023}$ & $6.8_{-2.9}^{+1.5}$ & $-4.25_{-2.46}^{+0.61}$ & $2.784$ \\
			& 3 & $0.441817_{-0.000015}^{+0.000016}$ & $53.9_{-7.3}^{+8.9}$ & $-1.00_{-0.13}^{+0.13}$ &  \\
			& 4 & $0.44100_{-0.00027}^{+0.00020}$ & $351_{-83}^{+131}$ & $-0.228_{-0.062}^{+0.061}$ &  \\
			\hline
			2014 August 03 & 1 & $0.4412917_{-0.0000033}^{+0.0000015}$ & $4.79_{-2.43}^{+0.82}$ & $-20.5_{-20.0}^{+2.7}$ & $2.858$ \\
			& 2 & $0.4412163_{-0.0000027}^{+0.0000027}$ & $8.1_{-1.1}^{+1.3}$ & $-4.64_{-0.63}^{+0.59}$ & $1.919$ \\
			\hline
			2014 September 01 & 1 & $0.4412914_{-0.0000012}^{+0.0000008}$ & $4.45_{-0.53}^{+0.30}$ & $-22.2_{-2.6}^{+1.2}$ & $2.815$ \\
			& 2 & $0.4412228_{-0.0000019}^{+0.0000023}$ & $6.7_{-2.7}^{+1.3}$ & $-4.55_{-2.42}^{+0.59}$ & $2.857$ \\
			& 3 & $0.441820_{-0.000017}^{+0.000017}$ & $55.7_{-7.8}^{+8.7}$ & $-0.82_{-0.11}^{+0.11}$ &  \\
			& 4 & $0.44050_{-0.00017}^{+0.00014}$ & $328_{-95}^{+119}$ & $-0.252_{-0.057}^{+0.049}$ &  \\
			\hline
			2014 Average & 1 & $0.44129230_{-0.00000041}^{+0.00000039}$ & $4.96_{-0.16}^{+0.15}$ & $-20.38_{-0.56}^{+0.51}$ & $2.945$\\
			& 2 & $0.4412209_{-0.0000011}^{+0.0000011}$ & $7.65_{-0.66}^{+0.64}$ & $-4.13_{-0.30}^{+0.27}$ & $2.583$\\
			& 3 & $0.441819_{-0.000010}^{+0.000010}$ & $54.2_{-5.0}^{+5.4}$ & $-0.900_{-0.075}^{+0.074}$ & \\
			& 4 & $0.44061_{-0.00014}^{+0.00014}$ & $338_{-64}^{+73}$ & $-0.197_{-0.031}^{+0.030}$ &  \\
			\hline
			2018 July 01 & 1 &  $0.4412719\pm0.0000038$ & $4.12\pm 0.13$ & $-19.41\pm0.55$&  \\ 
			& 2 & $0.4412030\pm0.0000038$ & $ 4.17\pm0.52$ & $-5.02\pm0.54$& \\
			\hline
	\end{tabular}}
	\begin{flushleft}
		\small
		\item \textbf{Notes.} Column 1 gives the observation epoch; column 2 the Gaussian component corresponding to that shown in Fig.~\ref{fig:spetrum} ; column 3 the component redshift; column 4 the component rest-frame FWHM; column 5 the peak component depth as a fraction of the continuum flux density; column 6 the pairwise peak position difference. Intervals of $1\sigma$ are given for the measured uncertainties, derived from Gaussian components fitting.
	\end{flushleft}
\end{table*}

Through this comparison, we must notice that a long-term consecutive observation with a consistent spectrum resolution is much necessary for a redshift drift measurement.   

\subsection{Observation plan for FAST}\label{sec:FAST}

The observable electromagnetic spectrum (70 MHz to 3 GHz) of FAST covers a wide range of redshifts in HI line observing. Using FAST, we plan to observe the SL effect through a combined observation mode (blind-searching plus targeted observation) in the coming years or decades. 

For the blind-searching mode which aims to discover as much HI absorption system candidates suitable for SL measurement as possible, we plan to join into the drift scan survey named CRAFTS which is specified in Ref.~\cite{Li:2018}, where multiple scientific goals coexist and force compromises wherever necessary. The CRAFTS observation will be made using the FAST L-band array of 19 feed horns (FLAN) and assembling at least four backends for pulsars, HI galaxies, HI imaging, and fast radio bursts (FRBs).

\begin{figure}
	\centering
	\includegraphics[width=0.8\textwidth]{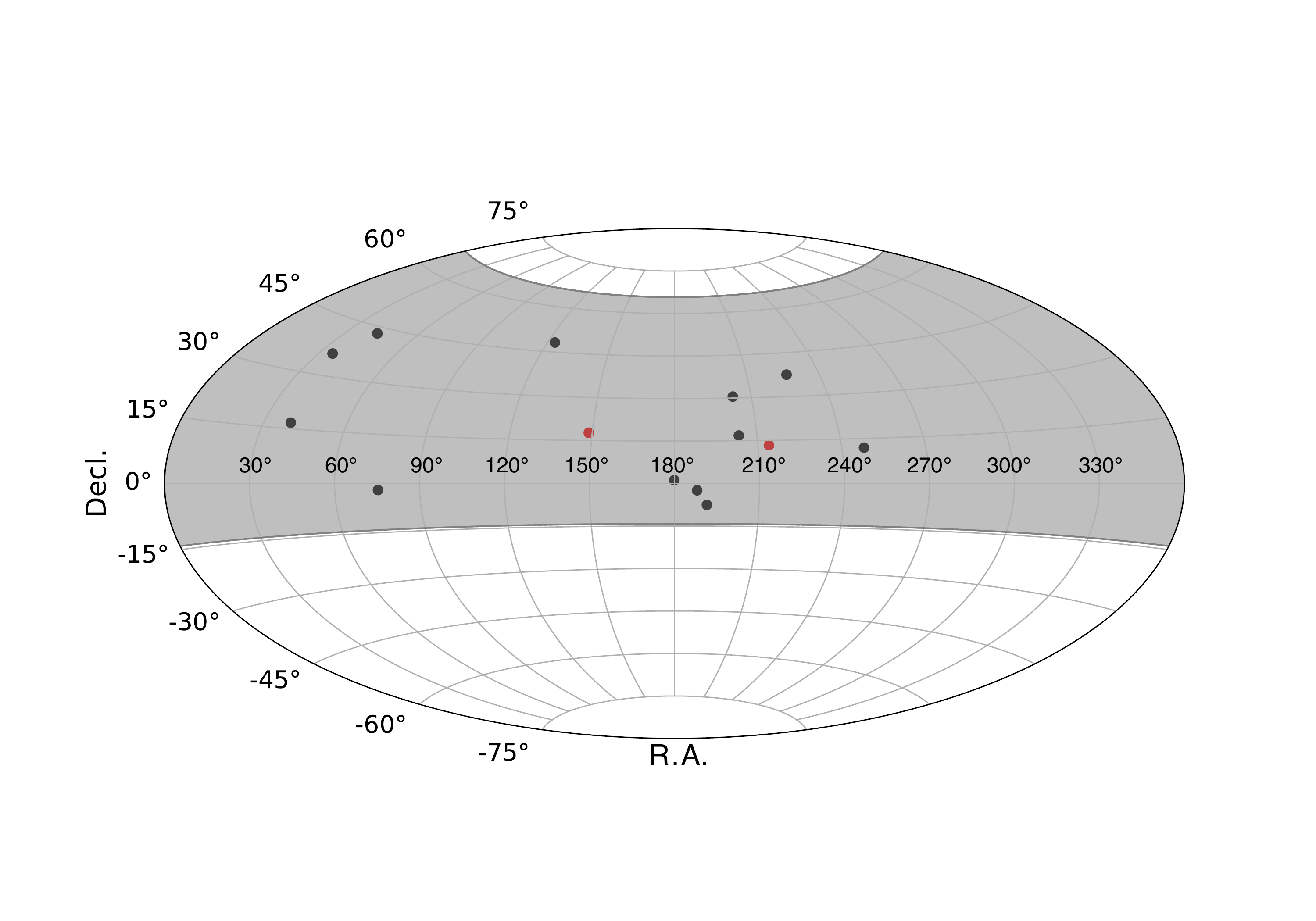}
	\caption{The shaded area in this aitoff projection of the equatorial celestial coordinates shows the sky coverage ($-14^\circ12^\prime < \rm DEC < 65^\circ48^\prime$) of CRAFTS survey, which covers $\Omega \simeq 7.272\ \rm sr$ ($57.9\%$ of the sky). These dots point out the sources (listed in Tab.~\ref{tab:sources}) which are selected as the first batch of candidates for the targeted observation mode, and the red two are covered within the FAST L band.}\label{fig:fov}
\end{figure}

The FAST 19-beam system achieves a better than $20\ \rm K$ receiver temperature over a 400 MHz band covering 1.04 -1.45 GHz corresponding to the redshift range $0<z_{\rm HI}<0.37$. With its high sensitivity ($A_{\rm eff}/T_{\rm sys}\simeq 1600-2000 \ \rm m^2/K$)\citep{Zhang:2019}, the sample size of the HI absorption systems will increase substantially. The 19-beam L-band focal plan array will be rotated to specific angles and receive continuous data streams, while the surface shape and the focal cabin stay fixed. FAST covers sky area $-14^\circ12^\prime < \rm DEC < 65^\circ48^\prime$, that is approximately 58\% of the sky (shown as Fig.~\ref{fig:fov}). The half power beam width HPBW $\sim 2.9^\prime$, with 19 beams total spacing $21.9^\prime$, such a survey will cover the northern sky in about 220 full days. Yu et  al. \cite{Yu:2017} predicts this highly efficient wide filed survey could be very competitive in blind-searching HI absorption systems. 

\begin{figure}
	\centering
	\includegraphics[width=0.8\textwidth]{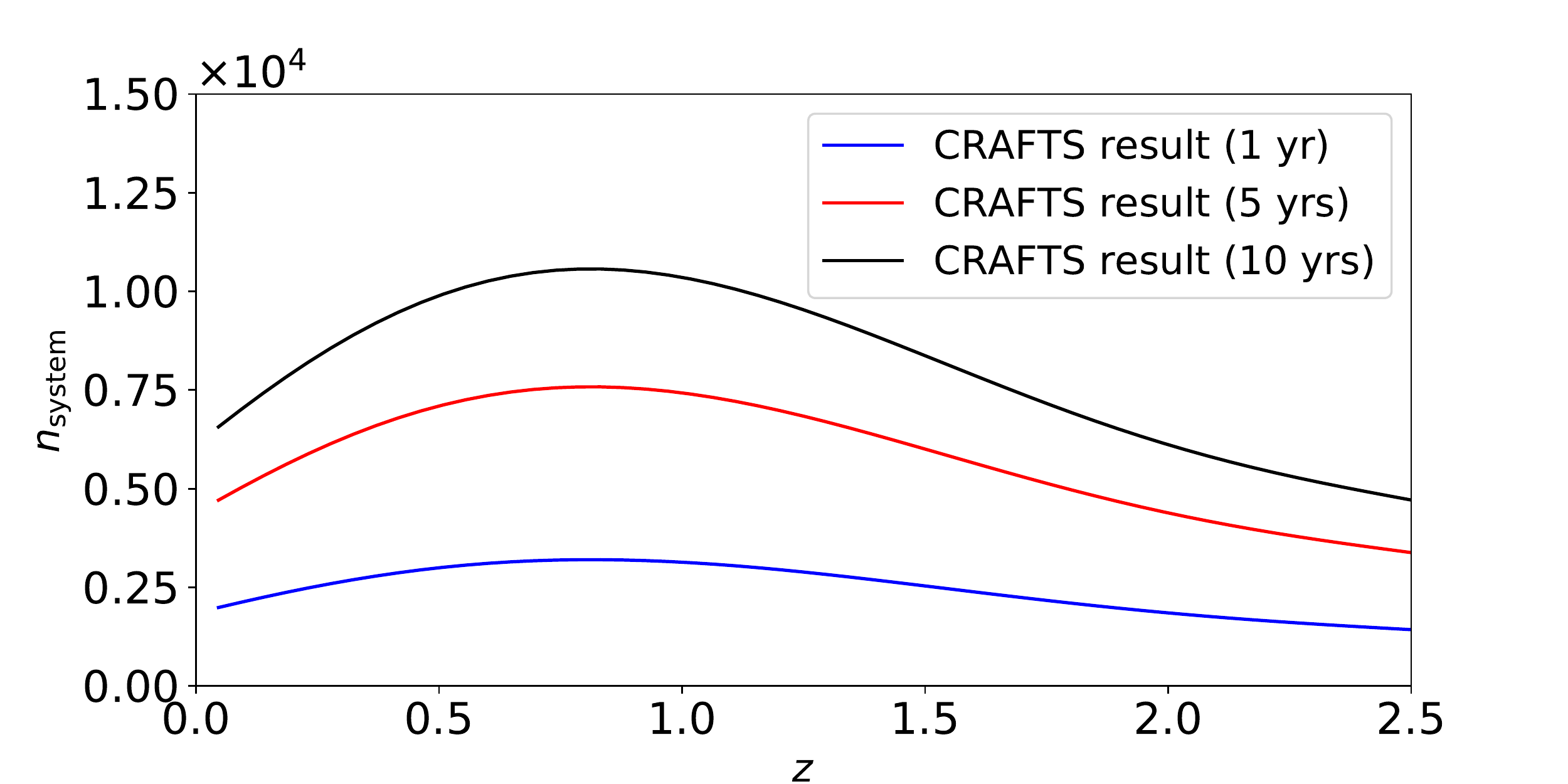}
	\caption{Forecasted observable HI absorption system's differential redshift distribution (${\rm d}N/{\rm d}z$)  of FAST sky survey in 1 year (bule), 5 years (red) and 10 years (black) respectively.}\label{fig:numbers}
\end{figure}

Here we evaluate the performance of CRAFTS for the blind-searching mode. The signal-to-noise ratio is estimated as the flux density $F$ of the observed source over it's measurement error $\Delta F$, for a dual polarized system
\begin{align}
S/N=F/\Delta F = F\sqrt{2\Delta \nu \Delta t}/\rm SEFD,
\end{align}
the system equivalent flux density ${\rm SEFD}=k_{\rm B}T_{\rm sys}/A_{\rm eff}\simeq 0.87\ \rm Jy$. For typical  equivalent width of absorber $u_{\rm width}=2\ \rm km/s$, the line width $\Delta \nu = u_{\rm width}\nu/c \sim 9 \rm kHz$, the integration time 
\begin{align}
\Delta t=\frac{\lambda_{\rm obs}}{2\pi D \cos(\delta)}t_{\rm scan},
\end{align}
where the observational wavelength $\lambda_{\rm obs} = 21\rm cm\times(1+z)$, the illuminated diameter $D=300\ \rm m $, $\delta$ denotes the declination and $t_{\rm scan}$ is the total observation time per scan strip respectively. A scan strip is a sky region defined as the solid angle  
\begin{align}
{\rm d}\Omega = 2\pi \cos(\delta)\rm d\delta,
\end{align} where $\rm d\delta$ is the beam offset. 
For greater than $10\sigma$ detections, the minimun source flux density $F_{\min} = 10\Delta F/r$, where the typical fractional depth of intervening 21-cm absorber is $r=20\%$. Therefore, we could calculate the differential redshift distribution for the observable absorption systems at a given declination $\delta$,
\begin{align}
n_{\rm system}(z,\delta)=n_{\rm DLA}(z){\rm d}\Omega\int_{F_{\min}}^{\infty}\int_{z}^{\infty}n_{\rm R}(z^\prime,F){\rm d}z^\prime{\rm d}F.
\label{eq:nsys}
\end{align}
The incidence of neutral gas DLAs at any given line of sight is a function of redshift for $0 < z < 5$ from an updated relation \citep{Rao:2017} as
\begin{align}
n_{\rm DLA}(z)=\frac{{\rm d}N_{\rm DLA}}{{\rm d}z} = (0.027 \pm 0.007)(1 + z)^{(1.682\pm 0.200)}.
\end{align}
We assume there exist an independence between the flux density distribution \citep{Condon:1984} and redshift distribution \citep{Zotti:2010}, thus 
\begin{align}
n_{\rm R}(z^\prime,F)=n_{\rm R}(z^\prime)f_{\rm R}(F).
\end{align}
Here we normalize the distribution function of differential source counts of radio sources as
\begin{align}
f_{\rm R}(F)=\frac{{\rm d}N_{\rm R}}{{\rm d}F}[\int_{10\rm mJy}^{F_{\rm inf}}\frac{{\rm d}n_{\rm R}}{{\rm d}F^\prime}{\rm d}F^\prime]^{-1}.
\end{align}
We set the lower limit of integration in normalization as $10 \rm mJy$, which is consistent with the lower limit flux density in the redshift distribution we applied. By integrating equation \ref{eq:nsys} over the sky coverage of FAST, we could estimate the redshift distribution of observable HI absorption systems shown as Fig.~\ref{fig:numbers}. Thanks for the wide field and high sensitivity of FAST, we predict that about $70$ systems will be discovered for one month scan around celestial euqator by the FAST L band receiver, and about $800, 1900$ and $2600 $ systems for 1 year, 5 years and 10 years CRAFTS survey respectively. Amounts of observable systems provide us a great sample to select candidates for targeted SL effect observations. 

\begin{figure}
	\centering
	\includegraphics[width=0.6\textwidth]{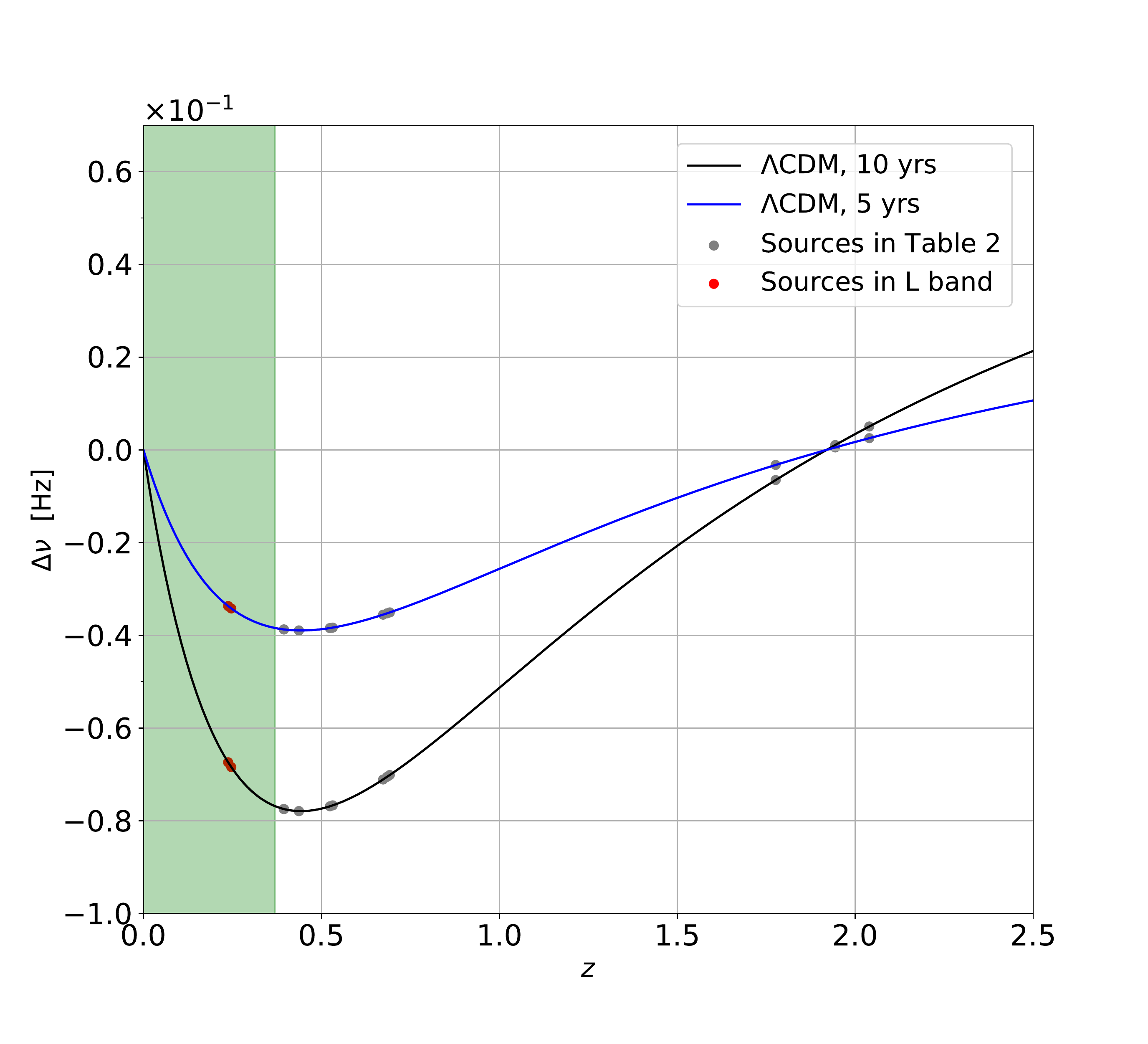}
	\caption{Theoretical frequency change indicates the requisite resolution for a ten-years (and a five-years) consecutive targeted mode to measure out the SL effect signal. The green shaded region shows the observable redshift range of the FAST 19-beam L band system. The dots on the lines mark out the required frequency resolution of the sources in Tab.~\ref{tab:sources}.}\label{fig:dotf}
\end{figure}

For targeted observation, we select out 14 sources within the sky coverage of FAST as the first batch of observation targets, which are shown in Table \ref{tab:sources}. The redshift drift will directly be quantified through the observed frequency change of the peak HI absorption line position
\begin{align}
	\Delta \nu = -\nu_{\rm e}\frac{\Delta z}{(1+z)^2},
\end{align}
where $\nu_{\rm e}$ is the rest frequency of HI absorption.
The required frequency resolution for a decadal (and five years) targeted mode to measure out the SL effect is shown as Fig.~\ref{fig:dotf}. At least sub-$0.1\rm Hz$ frequency resolution is required for the two sources (in red) covered in the L band. The shorter time span requires the higher spectral resolution. Apart from the listed sources, all newly discovered HI absorption systems through blind-searching should be added to this list. By comparing the spectral lines with different epochs (including the observation by \cite{Darling:2012}), we could analyze how stable of the spectral these candidates are and choose those with enough stability to measure out the $\dot{z}$ further.

\begin{table}
	\centering\caption{Selected sources for future SL measurement using FAST.}\label{tab:sources}
	\begin{tabular}{llll} 
		\hline
		Absorber & $z_{\rm HI}$ &	RA(J2000)  &	DEC(J2000)  \\
		&  &(decimal $\deg$) &	 ( decimal $\deg$) \\ \hline
		3C196 & 0.43667498(93)	&123.400138 &	48.217378\\	
		0248+430 &0.39408591(14) &	42.893903 &	43.254397	\\
		B3 1504+377 & 0.67324197(79) &	226.539708 &	37.514203	\\
		B0218+357 &	0.6846808(13)& 35.272792 &	35.937145	\\
		3C286 &	0.692153275(85) &202.784533 &	30.509155	\\
		\textbf{PKS 0952+179} & 0.2378155(16) &148.736765 &	17.725339\\	
		1331+170 &	1.7763904(66) & 203.399094 &	16.817782	\\
		0235+164 & 0.523741603(72) & 39.66209 &	16.616465	\\
		\textbf{PKS 1413+135} & 0.24670374(30) &	213.995072 &	13.33992\\	
		PKS 1629+120 & 0.5317935(11) &	247.938587 &	11.934165\\	
		1157+014 & 1.943670(12) & 179.936809 &	1.201982\\
		0458$-$020 & 2.0393767(21) &  75.303374 &	-1.987293	\\
		PKS 1229$-$021 & 0.39498824(59) &	188	& -2.4014	\\
		PKS 1243$-$072	&  0.4367410(14) &191.517633&	-7.512937\\	
		\hline
	\end{tabular}
	\begin{flushleft}
		\small
		\textbf{Notes.} The listed sources are HI absorption line systems selected from \cite{Darling:2012}, which are in the sky coverage of FAST and available for SL effect measurement. The absorbers in bold can be observed through the 19 beams L band receiver, while the ultral-wide-band receiver is needed for the others. The coordinates in the rest frame of RA/Dec Equatorial (J2000.0) are extracted from the Strasbourg astronomical Data Center (CDS)\footnote{http://cdsweb.u-strasbg.fr/}. 
	\end{flushleft}
\end{table}

\section{Conclusions and Discussions}\label{sec:conclu}

In this paper, we illustrate how the spectroscopic acceleration $\dot{v}_{\mathrm{spec}}$ is not suitable in quantifying the SL effect in high redshift. We also clarify the differences between the $\dot{v}_{\mathrm{spec}}$, the recession acceleration $\dot{v}_{\mathrm{rec}}$ and the cosmic acceleration $\ddot{a}$. As a result, we choose to use the observable $\dot{z}$ only. Then we report an approximately $3\ \rm km \ s^{-1}$ line shift for the peak position of HI absorption of PKS B1740-517 from our test observations using BETA in 2014 and Parkes in 2018. The shift falls into one channel spacing of frequency resolution of BETA, thus we conclude that it weakens any explanations in the source's peculiar acceleration and requires at least one more Parkes observation to test the observation consistency.  It also reminds us of the great significance for a consecutive consistent observation in measuring SL effect. 

We propose a combined observation mode for FAST, which is blind-searching for increasing the sample size of the HI absorption systems combined with long-term consecutive targeted observation to measure the spectral line shift. We forecast the performance of future CRAFTS survey in blind-searching within the sky coverage of FAST. Benefited from the successful deployment of the FAST 19-beam receiving system and its high sensitivity, we predict that about $800$ HI absorption systems will be detected for 1 year CRAFTS survey and up to $2600$ systems for 10 years. This high efficient survey will provide us amounts of candidate sources to be observed in the targeted mode thereby improve the statistics and precision of redshift drift measurement. The required frequency resolution for targeted observation is redshift related and inversely proportional to the observation time span. Overall, through 10 years of consecutive targeted spectroscopic observation with a level of sub-$0.1\rm Hz$ frequency resolution, we could detect the first-order derivative of the cosmological redshift. This precision is will help us distinguish cosmological models with a great significance.  

To fully use the coverable frequency of FAST, a lower band receiver is expected in CRAFTS survey for higher redshift observations. By combining with a southern sky survey using the telescopes such as Parkes or SKA \citep{Dewdney:2009,Kloeckner:2015}, we can greatly improve the accuracy and efficiency in SL cosmology. 

\section*{Acknowledgements}
We are grateful to Andrew Cameron, George Hobbs and Shi Dai for their kindness help in Parkes observation. We greatly thank James Allison and Maxim Voronkov for offering their observational data and helpful discussions. We thank anonymous referees very much for their comments on the appropriateness of the spectroscopic acceleration in SL effect representation.  We also thank Guojian Wang and  Yichao Li for useful comments and discussions. This work was supported by the National Key R \& D Program of China (2017YFA0402600), the National Science Foundation of China (Grants No. 11573006, 11528306, 11725313, 11690024), the Fundamental Research Funds for the Central Universities and the Special Program for Applied Research on Super Computation of the NSFC-Guangdong Joint Fund (the second phase) and by the CAS International Partnership Program NO.114A11KYSB20160008.

\bibliographystyle{JHEP}

\bibliography{sl_jcap}

\end{document}